\documentstyle[12pt]{article}

\begin{document}

\begin{center}

{\large \bf Topology Change in Canonical Quantum Cosmology} \\

\vspace{2cm}

Vitorio A. De Lorenci ${}^{(a)}$, J\'er\^ome Martin ${}^{(a),(b)}$, 
Nelson Pinto-Neto ${}^{(a)}$ and Ivano~Dami\~ao~Soares ${}^{(a)}$ \\
\vspace{1cm}
${}^{(a)}$ {\it Laborat\'orio de Cosmologia e F\'{\i}sica Experimental
de Altas Energias,} \\
\vspace{0.1cm}
{\it Centro Brasileiro de Pesquisas F\'{\i}sicas,} \\
\vspace{0.1cm}
{\it Rua Dr. Xavier Sigaud 150, Urca,} \\
\vspace{0.1cm}
{\it Rio de Janeiro CEP 22290-180-RJ, Brazil.} \\
\vspace{0.1cm} 
{\it E-mail: nen@lca1.drp.cbpf.br}\\
\vspace{0.7cm}
${}^{(b)}$ {\it Laboratoire de Gravitation et Cosmologies Relativistes,} \\
\vspace{0.1cm}
{\it Universit\'e Pierre et Marie Curie, CNRS/URA 769,} \\
\vspace{0.1cm}
{\it Tour 22/12, Boite courrier 142, 4 place Jussieu,} \\
\vspace{0.1cm}
{\it 75252 Paris, France.} \\
\vspace{0.1cm}
{\it E-mail: jmartin@ccr.jussieu.fr}

\vspace{1cm}

\begin{abstract}

\vspace{0.5cm}

We develop the canonical quantization of a midisuperspace model which 
contains, as a subspace, a minisuperspace constituted of a
Friedman-Lema\^{\i}tre-Robertson-Walker Universe filled with
homogeneous scalar and dust fields, where the sign of the intrinsic
curvature of the
spacelike hypersurfaces of homogeneity is not specified, allowing
the study of topology change in these hypersurfaces.
We solve the Wheeler-DeWitt equation of the midisuperspace model
restricted to this minisuperspace subspace in the semi-classical
approximation. Adopting the conditional probability interpretation,
we find that some of the solutions present change of topology of
the homogeneous hypersurfaces. However,
this result depends crucially on the interpretation we adopt:
using the usual probabilistic interpretation, we find selection 
rules which forbid some of these topology changes.   

\end{abstract}

\end{center}

\vspace{1cm}

Pacs numbers: 98.80.H, 98.80.K

\newpage
\section{Introduction}
One of the most interesting and fascinating aspects of Quantum Gravity is 
the idea that the topology of spacetime could change. This question was 
raised long ago by John Archibald Wheeler in his seminal papers on Quantum 
Gravity \cite{arch1}, where he described his ideas about the spacetime 
foam: quantum fluctuations of geometry could enhance quantum changes of
topology. In the classical domain, a well-known theorem due to Geroch 
\cite{ger} shows that a generic time orientable manifold endowed with 
a Lorentzian metric and exhibiting a topology change contains necessarily 
either closed timelike curves or singularities. If we further assume
Einstein's equations, topology changes are forbidden by the unique 
requirement that spacetime should remain non-singular \cite{tip}. 
Nevertheless, it exists some examples of Lorentzian manifolds presenting 
a topology change which have only mild singularities (the curvature 
invariants do not blow up at these points) \cite{hor0}. As these manifolds
have finite Einstein-Hilbert action, they should contribute to the
functional integral in the sum-over-histories approach to
Quantum Gravity. This fact suggests that topology changes could indeed
be possible quantum-mechanically. The aim of this paper is to describe 
some situations where classically forbidden changes of topology are indeed 
possible in the quantum domain.
\par
Usually, the articles discussing quantum changes of topology use the
path integral formalism and the semi-classical approximation. Some 
interesting results were obtained in Refs. \cite{col,str,hor} like, 
for instance, the proof that, due to the quantum tunneling effect, 
extremal black holes can be created in pair. Our approach will be based on 
canonical Quantum Gravity \cite{arch1,dew}, which assumes from the very 
beginning a topology of the type $R \times V^3$. Hence, only space 
topology changes in $V^3$ can be studied in this framework. In this 
paper, we will consider the following metric:
\begin{equation}
\label{defmetric}
{\rm d}s^2 = -N^2(t){\rm d}t^2 + a^2(t)\left\{ {\rm d}{\chi}^2 + 
\left[\frac{\sin(\sqrt{k(t)}\chi)}{\sqrt{k(t)}}\right]^2 
{{\rm d}\Omega}^2 (\theta ,\varphi)\right \},
\end{equation}
where ${{\rm d}\Omega}^2(\theta,\varphi)$ is the usual line element of a 
two-sphere. The metric (\ref{defmetric}) is a 
Friedman-Lema\^{\i}tre-Robertson-Walker (FLRW) like metric with a time 
dependent $k(t)$. The idea is that, when $k(t)$ passes through zero, a 
change of topology occurs \footnote{Note that, as the space volume 
depends on $k(t)$ and
$a(t)$, a small $k(t)$ does not necessarily implies a big space volume.}. 
\par
However, it is well-known that the metric tensor alone describes a manifold 
only locally, and does not provide information about its overall 
structure, i.e., about its topology. Therefore, it is necessary to 
discuss the features of the metric (\ref{defmetric}) in more details. 
In particular, we need to specify the ranges of the variables 
$(\chi,\theta,\varphi)$.
\par
One of the major tasks of the mathematical research of the nineteenth 
century was to elaborate a topological classification of spaces. For 
the purpose of illustration, let us first discuss the two-dimensional 
case since it is simple (it is sometimes possible to represent 
a two-dimensional space as a surface embedded in $E^3$), and since 
the complete classification is almost entirely known. One can show 
that any two-dimensional manifold $V^2$ is homeomorphic to the 
quotient $\tilde{V}^2/\Gamma$, where $\tilde{V}^2$ is the universal 
covering of $V^2$ and $\Gamma $ is the group of the covering 
transformations. The universal covering $\tilde{V}^2$ is either 
$R^2$ (the Euclidean plan), $S^2$ (the two-sphere) or $H^2$ 
(the hyperbolic plan). In 
addition the group $\Gamma $ is isomorphic to the fundamental group 
(or first homotopy group) of $V^2$: $\pi _1(V^2)$. Let us consider 
a simple example: the torus $T^2=S^1\times S^1$. Its universal 
covering is $R^2$ and its first homotopy group is $\pi _1(T^2)=
Z\oplus Z$. Thus, $T^2$ is homeomorphic to $R^2/(Z\oplus Z)$. Endowed 
with this theorem, we see that it is sufficient to know the possible 
groups $\Gamma $ for being able to classify the two-dimensional 
manifolds. The classification is complete (except when the universal 
covering is $H^2$ and the manifolds non compact), and can be found, for 
instance, in Ref. \cite{Coxeter}.
\par
Let us now consider a more restrictive situation where the surfaces under 
considerations are closed (here, we use the terminology {\it closed 
spaces } for a compact space without boundary). They are always 
homeomorphic to a polygon (called the {\it fundamental polyhedron}, 
which is nothing but the development of the largest simply-connected domain 
of the surface) whose edges are identified by pairs. In this case, 
one can show that the topology of a surface is uniquely specified 
by its orientability and the number of its holes (i.e., its genus). 
\par
Having at our disposal these mathematical results, we can now ask a 
slighty different question (but closer to the problem we are interested 
in): being given a two-dimensional metric, what are the  
different types of manifolds (i.e., with different topologies) that can 
be described locally by such a metric? For simplicity, let us consider 
the metric ${\rm d}s^2={\rm d}x^2+{\rm d}y^2$. Its scalar curvature 
is zero. Since, locally, a manifold and its universal covering are 
indistinguishable, the latter must be $R^2$. Then, we can use the 
previous classification to see that $R^2$, $R\times S^1$, the M\"obius band, 
$S^1\times S^1$ and the Klein bottle, all admit the metric under 
consideration. If, in addition, we require that the manifold be orientable 
and closed, the only possibility which remains is $T^2=S^1\times S^1$, as 
expected. Only the knowledge of some global property, like 
for instance the ranges of the variables $x$ and $y$ ($-L\leq x\leq L$, 
$-L\leq y\leq L$ for $T^2$, where the opposite edges of the square are  
identified and $-\infty \leq x\leq +\infty$, 
$-\infty \leq y\leq +\infty$ for $R^2$), can allow us to decide what is 
the manifold we deal with.
\par
On the other hand, we are obviously not obliged to write the flat metric 
in the cartesian form. We can also write it with the help of the polar 
coordinates: ${\rm d}s^2={\rm d}\chi ^2+\chi ^2{\rm d}\theta ^2$. Then, 
for the torus $T^2$ (we assume here that the fundamental polyhedron 
is a square whose edge has a length of $2L$), the ranges of the two 
variables are $0\leq \theta \leq 2\pi$ and $0\leq\chi\leq\chi _0(\theta)$,
where $\chi _0(\theta) = L/|\sin\theta|$ for $\pi/4\leq \theta \leq 3\pi/4$, 
$5\pi/4\leq \theta \leq 7\pi/4$ and  $\chi _0(\theta) = L/|\cos\theta|$ for
the other values of $\theta$. The opposite edges are still identified,
of course. We would like also to remark that $\chi _{max}\equiv 
\max\{\chi _0(\theta )\}=\sqrt{2}L $ is equal to the 
radius $\chi _{circ}$ of the circumscribed circle. 
As a consequence, the area of the torus 
is less than $\pi \chi _{max}^2=2\pi L^{2}$, 
the area of the circumscribed circle. Also, $\chi _{min}\equiv 
\min\{\chi _0(\theta )\}=L $ is equal to the 
radius $\chi _{insc}$ of the 
inscribed circle, and the area of the torus is greater than $\pi L^{2}$, the
area of the inscribed circle. 
Obviously, the value of $\chi _{insc}$, $\chi _{circ}$ depend on the 
size $L$ of the fundamental polyhedron.
\par
Let us now turn to the three-dimensional case. It actually corresponds 
to the real situation since we are interested in the topology of the 
spacelike sections $V^3$. More precisely, we would like to answer 
a question similar to the one already addressed before, namely: being 
given the metric of the three-dimensional hypersurfaces, 
\begin{equation}
\label{defhypersurf}
{\rm d}s^2 = a^{2}\left\{{\rm d}{\chi}^2 + 
\left[\frac{\sin(\sqrt{k}\chi)}{\sqrt{k}}\right]^2 
{{\rm d}\Omega}^2 (\theta ,\varphi)\right\},
\end{equation}
what are the topologies compatible with such a metric? The techniques 
described above are still applicable. The scalar curvature of this 
metric is constant and equal to $6k/a^2$. In addition, we will restrict 
our considerations to closed (and orientable) spaces in order to 
avoid possible surface terms in the Hamiltonian formalism. One can 
show that any compact three-manifold with constant curvature is 
homeomorphic to $\tilde{V}^3/\Gamma $, where the universal covering 
$\tilde{V}^3$ is either $R^3$, $S^3$ or $H^3$ according to the 
sign of $k$ ($k=0$, $k>0$ or $k<0$). The group $\Gamma $ is still the 
group of the covering transformations. In three dimensions the closed
space is now a polyhedron whose faces are identified by pairs. For a 
general review see Ref. \cite{luminet}. 
\par
Let us first treat the case of a vanishing curvature. It is possible 
to establish the following result: it exists six closed orientable 
spaces locally described by the metric (\ref{defhypersurf}) with 
$k=0$. Four of them admit a parallelepiped (in what follows, we 
will choose a cube) as the fundamental polyhedron, and the two remaining 
ones an hexagonal prism. Among the first category is the hyper torus 
$T^3=S^1\times S^1\times S^1$. For $T^3$, the ranges of the variables 
$(\chi,\theta ,\varphi )$ are given by:
\begin{equation}
\label{rangek=0}
0\leq \chi \leq \chi ^{(0)}_0(\theta ,\varphi) , \quad 
0\leq \theta \leq \pi, \quad 0\leq \varphi \leq 2\pi. 
\end{equation}
We will not need to know the explicit form of the function 
$\chi _0^{(0)}(\theta,\varphi )$. The maximum value of $\chi _0^{(0)}$, 
$\chi _{max}^{(0)}\equiv \max[\chi _0^{(0)}(\theta ,\varphi )]$, which is equal
to the radius $\chi_{circ}$ of the circumscribed sphere,
depends on the size of the fundamental cube, which is arbitrary,
 as it was the case in our two-dimensional 
example. This would be also true for any other three-manifold with 
$k=0$. In the remainder of this article, we will work exclusively with 
$T^3$ keeping however in mind that five other choices are possible.
\par
Let us now consider the case $k>0$. The universal covering is $S^3$. 
Since it is compact, all the three-spaces admitting this universal 
covering are also compact. There are an infinite number of such 
spaces \cite{seifert,Poincare}. We will consider two cases: $S^3$ itself 
and the Poincar\'e dodecahedral 
space, $D^3\equiv S^3/I^{*}$, where $I^{*}$ is the binary symmetry group of 
the icosahedron. The ranges of $(\chi,\theta,\varphi )$ can be now 
written as:
\begin{equation}
\label{rangek>0}
0\leq \chi \leq \chi ^{(1)}_0(\theta ,\varphi;V^3)/\sqrt{k} , \quad 
0\leq \theta \leq \pi, \quad 0\leq \varphi \leq 2\pi. 
\end{equation}
The symbol $V^3$ in the argument of $\chi _0^{(1)}$ indicates that 
the function $\chi ^{(1)}_0(\theta ,\varphi;V^3)$ is not the same 
whether $V^3$ is $S^3$ or $D^3$. We have 
$\chi ^{(1)}_0(\theta ,\varphi;S^3)=\pi $. This guarantees that no 
conical singularities will appear in 
this case. If $V^3=D^3$ the function $\chi _0^{(k=1)}$ is more 
complicated. In this case the maximum value of $\chi _0^{(1)}$ is not
arbitrary and is 
given by: $\chi _{max}^{(1)}\equiv \max[\chi _0^{(1)}(\theta, 
\varphi ;D^3)]\approx 0.163$.
\par
Finally, we consider the case $k<0$ where the universal covering is $H^3$. 
The classification of hyperbolic closed three-spaces is still an 
open question. Contrary to the $k=0$ case, the dimensions of the 
fundamental polyhedron are not arbitrary. Coxeter \cite{Coxeter} has 
described fifteen distinct possibilities obtained by tesselation of 
$H^3$. Among them, only four types 
have their fundamental polyhedron limited. In this article, we will 
consider the space $I^3\equiv H^3/\Gamma$ (where $\Gamma$ is an infinite
group) studied in Ref. \cite{fagundes}. 
The ranges of $(\chi,\theta,\varphi)$ are given by:
\begin{equation}
\label{rangek<0}
0\leq \chi \leq \chi ^{(-1)}_0(\theta ,\varphi)/\sqrt{-k} , \quad 
0\leq \theta \leq \pi, \quad 0\leq \varphi \leq 2\pi. 
\end{equation}
The maximum value of $\chi$ can be obtained from the radius of the 
circumscribed sphere (see above for a comparison with the two-dimensional 
case), $\chi _{circ}=\chi _{max}^{(-1)}
\equiv \max\{\chi ^{(-1)}_0(\theta ,\varphi)\}$, given by 
$\sinh \chi _{circ}=\sqrt{\tau ^6/4-1}\approx 1.867$ where 
$\tau \equiv (\sqrt{5}+1)/2$. Hence, we have 
$\chi _{max}^{(-1)}\approx 1.382$. The minimum value 
is given by the radius of the inscribed sphere 
$\chi _{insc}=\chi _{min}^{(-1)}\equiv \min
[\chi ^{(-1)}_0(\theta ,\varphi)]$, where $\sinh \chi _{insc}=
\sqrt{3\tau ^2/4-1}\approx 0.982$. This means that 
$\chi _{min}^{(-1)}\approx 0.868$.
\par
Let us now come back to the four-dimensional metric (\ref{defmetric}). 
Classically, if the matter fields are spatially homogeneous and isotropic, 
Einstein's equations yield the condition that $k(t)$ must be constant
in time. This comes from the momentum constraint equation, 
$G_{t\chi}=-T_{t\chi}$, which, in this case, is non-trivial giving 
$[\chi\dot{k}(t)]/[a(t)N(t)] = 0$ (a dot means derivation with respect to 
the time coordinate). Hence, a change of topology becomes 
impossible\footnote{Note that, due to the
high degree of symmetry of the metric and of the matter fields, the
requirements of regularity and causality are not 
necessary to forbid topology changes in this case. The momentum constraint
equation already do the job in a simple way.}. For more complicated matter 
fields having $T_{t\chi} \neq 0$, the spacetime presents a singularity
when $k(t)$ goes to zero. This is because the invariants contain terms 
like $\dot{k}(t)/k(t)$, which are divergent in the limit where $k(t)$ goes 
to zero, if $k(t)$ is regular in this vicinity. Our aim is to quantize 
the model based on Eq. (\ref{defmetric}) and to 
study whether or not a transition of the quantum observable 
associated with $k(t)$ is possible, i.e., if any of the wave 
functions found can describe, after having adopted a suitable 
interpretation, a spacetime exhibiting a change of topology. If so, 
this would demonstrate that 
topology changes on the hypersurfaces of homogeneity of a FLRW Universe 
can indeed occur when quantum effects become important.
\par
However, things are not that simple. It is not possible to construct
a {\it minisuperspace} Hamiltonian from the metric (\ref{defmetric}) because, 
as far as $k$ depends on time, this metric does not represent a 
spatially homogeneous spacetime, in the sense that the components of the 
four-dimensional curvature tensor in a local frame are functions of $t$ 
and $\chi$. The existence of the non-null component of the Einstein 
tensor $R_{t\chi}=[\chi\dot{k}(t)]/[a(t)N(t)]\neq0$ is a consequence of this
fact. Hence we are forced to introduce a {\it midisuperspace} model having 
a non-vanishing shift function $N_{\chi}(\chi, t)$. The metric, which was 
already proposed in Ref. \cite{pol} to study different 
problems, and also in Ref. \cite{Kief} to study quantum 
black holes, can now be written as:  
\begin{equation}
\label{22}
{\rm d}s^2 =\left[ -N^2(\chi,t) + \frac{N^{2}_{\chi}(\chi,t)}{a^{2}(\chi,t)}
\right] {\rm d}t^2 + 2 N_{\chi}(\chi,t) 
{\rm d}\chi \, {\rm d}t+ a^2(\chi,t)
[{\rm d}{\chi}^2 + {\sigma}^2(\chi,t) {{\rm d}\Omega}^2(\theta, \varphi)].
\end{equation}
The first step will be to carry out the quantization of this 
midisuperspace model. Then, in a second step, we will take into 
account in the quantum solutions the restrictions on the variables 
$a$ and $\sigma $, which from (\ref{22}), allow us to 
recover the metric (\ref{defmetric}). We see that 
consistency requires that we first treat the midisuperspace problem in 
order to come back to the minisuperspace model afterwards.

This article is organized as follows: in the next section we develop
the Lagrangian and Hamiltonian formalism for the midisuperspace model
defined by the metric (\ref{22}). The matter content of the model is 
described by a dust and a scalar fields $\xi(\chi,t)$ and $\phi(\chi,t)$. 
In section 3, the quantization of the model is achieved. The dust field is 
introduced to give a notion of time evolution to the quantum states \cite{21}.
We discuss the ordering and anomaly problems, and present some sets 
of semi-classical solutions to the Wheeler-DeWitt equation coming
from metric (\ref{22}), restricted to the subspace of the 
metric (\ref{defmetric}) and the 
homogeneous fields $\xi(t)$, $\phi(t)$. In section 4, we address the issue 
of interpretation. The dust field permits us to
adopt the usual probabilistic interpretation to the wave function of the 
Universe. We will show that, like a kind of ``selection rule", a topology 
change from $k=1$ and $S^{3}$ topology, to other values of $k$ is 
not allowed simply because the requirement of
normalization of our solutions implies that they must vanish when the 
topology of $V^3$ is $S^{3}$. On the other hand, adopting the 
conditional probability interpretation, 
which does not require normalizable wave functions, we will demonstrate 
that topology changes become possible in this framework even in the
case of the $S^{3}$ space. In section 5, 
we conclude with comments and perspectives for future works. Finally, 
the appendix presents other solutions to the quantum equations
which, however, are not suitable for the analysis of topology change.
We use units such that $c=1$ and $\kappa/2\pi = 1$, where $\kappa$ 
is the Einstein\rq s constant.

\section{Hamiltonian Formalism for the Classical Model}
\label{The Classical Model}

In the Arnowitt-Deser-Misner formalism \cite{ADM} the Einstein-Hilbert
Lagrangean density is given by:
\begin{equation}
{\cal L}_{g}[N,N_{i},\gamma_{ij}] = 
\frac{N\gamma^{\frac{1}{2}}}{4\pi}\left(\mbox{}^{3}\hspace{-.1cm}R 
+ K^{ij}K_{ij} - K^{2}\right),
\label{EH}
\end{equation}
where dynamically irrelevant total derivatives have been dropped. In the 
above expression $\gamma_{ij}$ is the metric of 3-dimentional spatial sections
$t=\mbox{constant}$. $K_{ij}$ is the extrinsic curvature, $K$ its trace and 
$\mbox{}^{3}\hspace{-.1cm}R$ is the scalar curvature of these sections.
$N$ and $N_{i}$ are the lapse and shift functions. 
It will be assumed that the matter content of the model is 
a minimally coupled scalar field $\phi(\chi,t)$  whose Lagrangian 
density can be written as:
\begin{equation}
{\cal L}_{\phi}[\phi] = -\frac{\sqrt{-g}}{4\pi}\left[\frac{1}{2}g^{\alpha\beta}
\phi_{,\alpha}\phi_{,\beta} + U(\phi)\right],
\label{scalar field}
\end{equation}
and a distribution of irrotational dust particles described by a dust 
field, see Refs. \cite{21,sal1}, with Lagrangian density defined by:
\begin{equation}
 {\cal L}_{\xi} = -\frac{\sqrt{-g}}{4\pi}\left[\frac{n}{2m}
\left(g^{\alpha\beta}\xi_{,\alpha}\xi_{,\beta} + m^{2}\right) 
+ V(\xi)\right],
\label{dust field}
\end{equation}
where $m$ is the mass, and $n$ is the rest number 
density of the dust particles.
The dust field $\xi(\chi,t)$ defines the 4-velocity field for dust
particles given by
\begin{equation}
U^{\mu}=-\frac{g^{\mu\nu}\xi_{,\nu}}{m},
\end{equation}
so that $\xi = \mbox{constant}$ will determine a congruence of space-like
hypersurfaces foliating the space-time. Therefore, this dust field may 
be used as the time variable for our model, and in the Schr\"odinger type 
equation obtained from the 
Wheeler-DeWitt equation. Indeed the rest number density $n$
appearing in Eq. (\ref{dust field}) has its canonically conjugated momentum
$\pi_{n}\approx 0$. Maintaining this constraint demands that 
$\dot{\xi} = N$ whenever $\xi^{'} = 0$.
 
For the class of geometries described by Eq. (\ref{22}), we define the 
total Lagrangian density by the expression,
\begin{equation}
\label{denslagtot}
{\cal L}={\cal L}_g+{\cal L}_{\phi }+{\cal L}_{\xi },
\end{equation}
which takes the following form:
\begin{eqnarray}
{\cal L} & = & \frac{a^{3}\sigma^{2}\sin\theta}{4\pi N}\left\{
- 6\left(\frac{\dot{a}}{a}\right)^{2} 
- 8\left(\frac{\dot{a}\dot{\sigma}}{a\sigma}\right)
- 2\left(\frac{\dot{\sigma}}{\sigma}\right)^{2}
+ 4\frac{\dot{a}}{a^{2}}\left[\left(\frac{N_{\chi}}{a}\right)^{'}
+ \frac{2N_{\chi}}{a}\left(\frac{a^{'}}{a} 
+ \frac{\sigma^{'}}{\sigma}\right)\right]
\right.
\nonumber\\
& & +\frac{4\dot{\sigma}}{a\sigma}\left[\left(\frac{N_{\chi}}{a}\right)^{'}
+ \frac{N_{\chi}}{a}\left(\frac{a^{'}}{a} 
+ \frac{\sigma^{'}}{\sigma}\right)\right]
- \frac{2N_{\chi}^{2}}{a^{4}}\left(\frac{a^{'}}{a}+\frac{\sigma^{'}}
{\sigma}\right)^{2} - 
\frac{4N_{\chi}}{a^{3}}\left(\frac{N_{\chi}}{a}\right)^{'}\left(\frac{a^{'}}{a}
+ \frac{\sigma^{'}}{\sigma}\right)
\nonumber\\
& & +\frac{\dot{\phi}^{2}}{2} - \frac{N_{\chi}\dot{\phi}\phi^{'}}{a^{2}} 
- N^{2}\left[\frac{{\phi^{'}}^{2}}{2a^{2}} + U(\phi) + V(\xi)\right] 
+ \frac{N_{\chi}^{2}}{2a^{4}}{\phi^{'}}^{2} 
\nonumber\\
& & \left.+\frac{n}{2m}\left[\dot{\xi}^{2} 
- \frac{2N_{\chi}}{a^{2}}\dot{\xi}\xi^{'}
- N^{2}\left(\frac{{\xi^{'}}^{2}}{a^{2}}+ m^2\right) + 
\frac{N_{\chi}^{2}\xi^{'2}}{a^{4}}\right]\right\},
\label{totalL}
\end{eqnarray}
where a dot and a prime denote derivative with respect
to $t$ and $\chi$, respectively. The total Lagrangian of our model is
given by:
\begin{equation}
\label{FLUZAO}
L = \int_{V^{3}}{\rm d}\chi{\rm d}\theta{\rm d}\varphi {\cal L}(\chi,t)
\end{equation}
If 
$\chi _0(\theta ,\varphi )$ does not depend on ($\theta $, 
$\varphi $) (see the discussion above) then the integration of the 
Lagrangian with respect to these variables can be performed. This 
amounts to replacing $\sin\theta$ with $4\pi$ in the expression given by Eq. (\ref{totalL}). In that case, it  will only remain a single integral over $\chi $ in 
Eq. (\ref{FLUZAO}). Note, however, that for
the cases where $\chi_{0}(\theta,\varphi)$ actually depends on 
($\theta,\phi$), the integration over $\theta$ and $\varphi$ cannot be done
independently of the integration over $\chi$. The triple integral  
in Eq. (\ref{FLUZAO}) will remain.

Let us now describe the Hamiltonian formalism. We begin with the simplest
case where the domain of integration in $\chi$ does not depend on $\theta$ and
$\varphi$.    
The canonical momenta are defined by 
\begin{eqnarray}
\pi_{Q_{i}} \equiv \frac{\delta L}{\delta\dot{Q_{i}}},
& \mbox{for} & Q_{i} \equiv \{a,\sigma,\phi,\xi\}.
\end{eqnarray}
Using the Lagrangian $L$ given by Eq. (\ref{FLUZAO}) and
(\ref{totalL}), and performing the integration on the $\theta$ and $\varphi$,
we easily find that: 
\begin{eqnarray}
\pi_{a} & = & \frac{4}{N}\bigl(-3a\dot{a}\sigma^{2}
-2a^{2}\sigma \dot{\sigma}+ 2\sigma \sigma 'N_{\chi} 
+\sigma ^2N_{\chi }'\bigr),
\label{pia} \\
\pi_{\sigma} & = & 
\frac{4a}{N\sigma }\bigl(-2a\dot{a}\sigma-a^{2}\dot{\sigma}
+\sigma^{'}N_{\chi }+\sigma N_{\chi}^{'} \bigr),
\label{pisigma}\\
\pi_{\phi} & = & \frac{a^{3}\sigma^{2}}{N}\bigl(\dot{\phi}
- \frac{N_{\chi}}{a^{2}}\phi^{'}\bigr),
\label{piphi}\\
\pi_{\xi} & = & \frac{na^{3}\sigma^{2}}{mN}\bigl(\dot{\xi}
- \frac{N_{\chi}\xi^{'}}{a^{2}}\bigr).
\label{pichi}
\end{eqnarray}
The total Hamiltonian density can be calculated by performing the 
Legendre transformation ${\cal H} = \sum_{i}\dot{Q_{i}}\pi_{Q_{i}} 
- {\cal L}$, 
yielding the following expression:
\begin{equation}
{\cal H} = N{\cal H}^{0} + N_{\chi}{\cal H}^{\chi},
\label{totalH}
\end{equation}
where ${\cal H}^{0}$ and ${\cal H}^{\chi}$ are the 
super-Hamiltonian and super-momentum constraints respectively, and are given
by the following equations:
\begin{eqnarray}
\label{H0}
{\cal H}^{0} & = & \left[\frac{\pi_{a}^{2}}{8a\sigma^{2}} 
+ \frac{3\pi_{\sigma}^{2}}{8a^{3}} 
- \frac{\pi_{a}\pi_{\sigma}}{2a^{2}\sigma} 
+ \frac{\pi_{\phi}^{2}}{2a^{3}\sigma^{2}}\right]
+ \sigma^{2}a^{3}{\cal V}
+ \sqrt{\frac{{\xi^{'}}^{2}}{a^{2}} + m^2}\,\,\pi_{\xi}, \\
\label{Hchi}
{\cal H}^{\chi} & = & \frac{1}{a^{2}}\left(- a\pi_{a}^{'} 
+ \sigma\pi_{\sigma}^{'} 
+ 2\sigma^{'}\pi_{\sigma} + \phi^{'}\pi_{\phi} + \xi^{'}\pi_{\xi}\right),
\end{eqnarray}
the ``superpotential'' ${\cal V}$ being defined  by the formula:
\begin{equation}
{\cal V} = -\,\mbox{}^{3}\hspace{-.1cm}R 
+ \frac{{\phi^{'}}^{2}}{2a^{2}} + U(\phi) + V(\xi).
\label{potencialas}
\end{equation}

Here it is worth remarking that Eqs. (\ref{totalH})-(\ref{Hchi}) 
can be derived directly from the Hamiltonian 
${\cal H} = \dot{\gamma}_{ij}\pi^{ij} + \dot{\phi}\pi_{\phi} +
\dot{\xi}\pi_{\xi} - {\cal L}$, by calculating 
\begin{equation}
\pi_{ij} = \frac{\delta{\cal L}}
{\delta\dot{\gamma}_{ij}} = \frac{\partial{\cal L}}{\partial\dot{a}}
\frac{\delta\dot{a}}{\delta\dot{\gamma}_{ij}} + 
\frac{\partial{\cal L}}{\partial\dot{\sigma}}
\frac{\delta\dot{\sigma}}{\delta\dot{\gamma}_{ij}}.
\end{equation}
for the class of geometries described by Eq. (\ref{22}). 

Also, the Poisson
Brackets (PB) for the variables ($Q_{i},\pi_{Q_{i}}$), are given by
\begin{equation}
\{Q_{i}(\chi,t),\pi_{Q_{j}}(\bar{\chi},t)\} =
\delta_{ij}\delta(\chi-\bar{\chi}),
\label{PB}
\end{equation}
all others being zero. We can check that 
the PB (\ref{PB}) are consistently derived from the standard
PB of the ADM variables ($\gamma_{ij},\pi^{ij}$), and that the Hamiltonian 
(\ref{totalH}) provides the correct Einstein\rq s equations for 
our model.

Before proceeding in the quantization of the above system, 
it will prove useful to perform a change to new variables
$\alpha(\chi,t)$ and $\beta(\chi,t)$ defined by
\begin{equation}
\alpha = \ln{a},
\hspace{1cm}
\beta  =- 2\ln(\sigma a).
\end{equation}
The momenta $\pi_{\alpha}$ and $\pi_{\beta}$ are related to the old 
momenta $\pi_{a}$ and $\pi_{\sigma}$ by 
\begin{eqnarray}
\pi_{a} & = & e^{-\alpha}\left(\pi_{\alpha}-2\pi_{\beta}\right),\\
\pi_{\sigma} & = &-2e^{\alpha+\frac{\beta}{2}}\pi_{\beta}.
\end{eqnarray}

In the new variables $\alpha$ and $\beta$,
the super-Hamiltonian and super-momentum constraints can be expressed as
\begin{eqnarray}
\label{super-hamiltonian}
{\cal H}^{0} & =& e^{-\alpha+\beta}
\left(\frac{\pi_{\alpha}^{2}}{8} 
+ \frac{\pi_{\alpha}\pi_{\beta}}{2} 
+ \frac{\pi_{\phi}^{2}}{2}\right)
+ e^{\alpha-\beta}{\cal V}
+ \sqrt{e^{-2\alpha}{\xi^{'}}^{2} 
+ m^2}\,\,\pi_{\xi}, \\
\label{super-momentum}
{\cal H}^{\chi} &=& e^{-2\alpha}\left(\alpha^{'}\pi_{\alpha}
+ \beta^{'}\pi_{\beta} + \phi^{'}\pi_{\phi} + \xi^{'}\pi_{\xi}
-\pi_{\alpha}^{'}\right),
\end{eqnarray}
where the functions ${\cal V}$ and $\,\mbox{}^{3}\hspace{-.1cm}R$ 
in the ${\alpha,\beta}$ variables have now the form:
\begin{eqnarray}
{\cal V} &=& -\,\mbox{}^{3}\hspace{-.1cm}R 
+ e^{-2\alpha}\frac{{\phi^{'}}^{2}}{2} + U(\phi) + V(\xi),  \\
\label{courbure}
\mbox{}^{3}\hspace{-.1cm}R &=& e^{-2\alpha}\left(
- \frac{3}{2}{\beta^{'}}^{2} - 2\alpha^{'}\beta^{'} 
+ 2\beta^{''} + 2e^{2\alpha + \beta}\right).
\end{eqnarray}
Finally, for further reference we also introduce the two variables 
$u(\chi,t)$ and $v(\chi,t)$ given by:
\begin{eqnarray}
u \equiv \ln\sigma = -\alpha - \frac{\beta}{2} ,\qquad 
v  \equiv \ln(a^{3}\sigma^{2}) = \alpha - \beta.
\end{eqnarray}
We remark that $v$ is nothing but the logarithm of the volume density
$\gamma^{\frac{1}{2}}$ divided by $\sin\theta$.
In these variables the super-Hamiltonian constraint is expressed as:
\begin{eqnarray}
{\cal H}^0 &=& e^{-v}\left[\frac{3}{8}\left(\pi^{2}_u - \pi^{2}_v\right)
+\frac{1}{2}\pi^{2}_{\phi }\right] + 
\sqrt{e^{\frac{4u}{3}-\frac{2v}{3}}\xi^{'2} 
+ m^2}\,\,\pi_{\xi} \nonumber \\
& & + e^{v}\left[-\mbox{}^{3}\hspace{-.1cm}R + e^{\frac{4u}{3}
+ \frac{2v}{3}}\frac{\phi ^{'2}}{2} + U(\phi) + V(\xi)\right], 
\label{HWDW2J}
\end{eqnarray}
exhibiting explicitly the Klein-Gordon character of the gravitational sector.

For the cases where $0\leq\chi\leq\chi_{0}(\theta,\varphi)$, the integration
on the $(\theta,\varphi)$ variables cannot be done as previously. In these
cases, the Hamiltonian which yields the Einstein\rq s  equations is
\begin{eqnarray}
\label{332}
H = \frac{1}{4\pi}\int_{0}^{2\pi}{\rm d}\varphi\int_{0}^{\pi}{\rm d}\theta
\sin\theta
\int_{0}^{\chi_{0}(\theta,\varphi)}{\rm d}\chi\left[N(\chi,t)
{\cal H}^{0}(\chi,t) + N_{\chi}(\chi,t)
{\cal H}^{\chi}(\chi,t) \right],
\end{eqnarray}
where ${\cal H}^{0}(\chi,t)$ and ${\cal H}^{\chi}(\chi,t)$ are
exactly the same as the ones given in Eqs. (\ref{H0}) and (\ref{Hchi}), or in
Eqs. (\ref{super-hamiltonian}) and 
(\ref{super-momentum}). The momenta also given by 
Eqs. (\ref{pia})-(\ref{pichi}) and the Poisson brackets are the same as
(\ref{PB}). The constraints are still given by:
\begin{eqnarray}
\frac{\delta H}{\delta N(\chi,t)} = 0 
&\Rightarrow 
& {\cal H}^{0}(\chi,t) = 0,
\label{34}\\
\frac{\delta H}{\delta N_{\chi}(\chi,t)} = 0 
&\Rightarrow 
& {\cal H}^{\chi}(\chi,t) = 0.
\label{35}
\end{eqnarray}

We would like to emphasize that the Hamiltonian (\ref{332}) furnishes
the correct Einstein\rq s equations for {\it whatever} function
$\chi_{0}(\theta,\varphi)$\footnote{The Hamiltonian density in Eq. (\ref{332}) 
may be obtained
directly from the Langrangian ``density" 
$\bar{{\cal L}} = 4\pi{\cal L}/\sin\theta$. The
quantity $\bar{{\cal L}}$ is still a density with respect to transformations
involving only the $\chi$ variable. One should remember that the fields depend only 
on $\chi$.}.

Let us now turn to the quantization of the classical model 
described above.

\section{Quantization}
\label{Quantization}

In order to quantize the constrained system described in the last 
section, we follow the Dirac\rq s prescriptions \cite{Dirac}. The variables of 
the phase space go into operators acting on the wave 
functional $\Psi[a,\sigma,\phi,\xi]$, the Poisson brackets turn into 
commutators and the momentum operators are represented by:
\begin{eqnarray}
\hat{\pi}_{Q_{i}}  \equiv  -i\hbar\frac{\delta}{\delta Q_{i}},
& \mbox{for} & Q_{i} \equiv \{a,\sigma,\phi,\xi\}.
\label{momenta}
\end{eqnarray}
Also, the space of the wave functionals must be endowed with the structure 
of a Hilbert space in order to associate operators with observables. 
Finally, the super-Hamiltonian and super-momentum constraints given by 
Eqs. (\ref{super-hamiltonian}), (\ref{super-momentum}), (\ref{34}) and
(\ref{35}) result in two 
relations which express that only a restricted region of the Hilbert 
space of the wave functionals contains the physical states of the 
theory:
\begin{equation}
\label{wdw-mc}
\hat{{\cal H}}^{0} \Psi[a,\sigma,\phi,\xi] = 0,\qquad
\hat{{\cal H}}^{\chi} \Psi[a,\sigma,\phi,\xi] =0.
\end{equation}
The first equation is the well-known Wheeler-DeWitt equation, whereas  
the second one is the so-called momentum constraint equation. In the 
remainder of this paper, we will be mainly interested in solutions of 
these equations. Note that the formulas (\ref{wdw-mc}) are valid for
all possible domains of variation of $\chi$, and thus applicable to all
possible topologies we may consider.
\par
At this point, it should be emphasized that the knowledge of the wave 
functional alone does not allow us to compute probabilities and/or mean 
values unless a measure has been specified. In order to address this 
question we notice that the Wheeler-De Witt Hamiltonian (\ref{HWDW2J}) 
with the following factor ordering:
\begin{eqnarray}
\label{HWDW1}
\hat{{\cal H}}^{0} &=& \frac{3}{8}[\hat{\pi}_{u}e^{-v}\hat{\pi}_{u} - 
\hat{\pi}_{v}e^{-v}\hat{\pi}_{v}]
+ \frac{1}{2}\hat{\pi} _{\phi }e^{-v}
\hat{\pi} _{\phi }+\sqrt{m^2 + e^{\frac{2}{3}\left(2u-v\right)}
\xi ^{'2}}\,\,\hat{\pi} _{\xi } 
\nonumber \\
& & + e^{v}[-\mbox{}^{3}\hspace{-.1cm}R + \frac{1}{2}e^{\frac{2}{3}
\left(2u-v\right)}\phi ^{'2} + U(\phi) + V(\xi)] \\
\label{HWDW2}
&=& \frac{3}{8}e^{-v}[\hat{\pi }_u^2-\hat{\pi}_v^2 - 
i\hbar\hat{\pi} _v \delta (0)]
+ \frac{e^{-v}}{2}\hat{\pi }_{\phi }^2 + \sqrt{m^2 
+ e^{\frac{2}{3}\left(2u-v\right)}
\xi ^{'2}}\,\,\hat{\pi }_{\xi } \nonumber \\
& & + e^{v}[-\mbox{}^{3}\hspace{-.1cm}R + 
\frac{e^{\frac{2}{3}\left(2u-v\right)}}
{2}\phi ^{'2} + V(\phi) + U(\chi)],
\end{eqnarray}
is hermitian if one chooses the ``cartesian" measure defined by: 
\begin{equation}
\label{defmeasure}
{\cal D}u{\cal D}v{\cal D}\phi.
\end{equation}
Note that $\hat{\pi}_{\xi}$ appears linearly in the Hamiltonian constraint
(\ref{HWDW2}) allowing to interpret the dust field $\xi$ as a time 
variable at the quantum level. This is why ${\cal D}\xi$ does not 
appear in the measure (\ref{defmeasure}).
In the remainder of this article we will adopt this measure.
\par
We would like to stress that the anomaly problem is still present. Indeed, 
strictly speaking, we should check that the algebra of the constraints 
is preserved at the quantum level. However, in the following, we will be 
dealing exclusively with WKB solutions valid at the $\hbar ^0$ order and 
this problem will not be important for us. Also the solutions will be 
independent of the factor ordering chosen previously. 

The super-Hamiltonian 
given by Eq. (\ref{HWDW2J}) can be re-expressed in terms of the variables 
$\alpha $ and $\beta $. We 
obtain the following expression for $\hat{{\cal H}}^{0}$:
\begin{eqnarray}
\hat{{\cal H}}^{0} & = &
\left[\frac{\hat{\pi}_{\alpha}e^{-\alpha+\beta}\hat{\pi}_{\alpha}}{8} 
+ \frac{1}{4}\left(\hat{\pi}_{\alpha}e^{-\alpha+\beta}\hat{\pi}_{\beta} 
+ \hat{\pi}_{\beta}e^{-\alpha+\beta}\hat{\pi}_{\alpha}\right) 
+ \frac{\hat{\pi}_{\phi}^{2}}{2}\right] 
+\,e^{\alpha-\beta}{\cal V} \nonumber \\
& & + \sqrt{e^{-2\alpha}{\xi^{'}}^{2} + m^2}\,\,\hat{\pi}_{\xi}.
\label{super-hamiltonian-ordering}
\end{eqnarray}
As announced previously, the equation $\hat{{\cal H}}^{0}\Psi = 0$ is a 
Schr\"odinger like functional equation with $\xi$ playing the role of 
time. This Hamiltonian is obviously hermitian in the cartesian 
measure ${\cal D}\alpha{\cal D}\beta{\cal D}\phi$ and can be also 
re-written as:
\begin{eqnarray}
\hat{{\cal H}}^{0} & = & e^{-\alpha+\beta}
\left[\frac{\hat{\pi}_{\alpha}^{2}}{8} 
+ \frac{\hat{\pi}_{\alpha}\hat{\pi}_{\beta}}{2} 
+ \frac{\hat{\pi}_{\phi}^{2}}{2}\right] 
+ i\hbar e^{-\alpha+\beta}\left(\frac{3\hat{\pi}_{\alpha}}{8}
- \frac{\hat{\pi}_{\beta}}{4}\right)\delta(0) 
+\, e^{\alpha-\beta}{\cal V} \nonumber \\
& & + \sqrt{e^{-2\alpha}{\xi^{'}}^{2} + m^2}\,\,\hat{\pi}_{\xi}.
\label{super-hamiltonian-anomaly}
\end{eqnarray}
\par
Let us now turn to the solutions of Eqs. (\ref{wdw-mc}). A general wave 
functional can always be put under the following form:
\begin{equation}
\label{psipol}
\Psi[\alpha,\beta,\phi,\xi] = 
e^{\frac{i}{\hbar}{\cal S}[\alpha,\beta,\phi,\xi]},
\end{equation}
where the functional ${\cal S}[\alpha,\beta,\phi,\xi]$ can be 
complex. If we introduce the form (\ref{psipol}) into the Wheeler-De Witt 
equation and the momentum constraint, we find that ${\cal S}$ must 
satisfy the equations: 
\begin{eqnarray}
\label{WDWS}
-i\hbar e^{-\alpha+\beta}
\left[\frac{1}{8}\frac{\delta^{2}{\cal S}}{\delta\alpha^{2}}
+ \frac{1}{2}\frac{\delta^{2}{\cal S}}{\delta\alpha\delta\beta}
+ \frac{1}{2}\frac{\delta^{2}{\cal S}}{\delta\phi^{2}}
- \left(\frac{3}{8}\frac{\delta{\cal S}}{\delta\alpha} 
- \frac{1}{4}\frac{\delta{\cal S}}{\delta\beta}\right)\delta(0)\right] + & & 
\nonumber\\
\hspace{-4mm}e^{-\alpha+\beta}\left[\frac{1}{8}
\left(\frac{\delta{\cal S}}{\delta\alpha}\right)^{2}\hspace{-2mm}+\hspace{-0.5mm}
\frac{1}{2}\left(\frac{\delta{\cal S}}{\delta\alpha}\right)\hspace{-2mm}
\left(\frac{\delta{\cal S}}{\delta\beta}\right)
\hspace{-0.5mm}+\hspace{-0.5mm}\frac{1}{2}\left(\frac{\delta{\cal S}}{\delta\phi}\right)^{2}\right] 
+ e^{\alpha-\beta}{\cal V} + \sqrt{e^{-2\alpha}{\xi^{'}}^{2} 
\hspace{-0.5mm}+\hspace{-0.5mm}m^2}\frac{\delta{\cal S}}{\delta\xi} & = & 0, \\
\label{momentumS}
e^{-2\alpha}\left[\alpha^{'}\frac{\delta{\cal S}}{\delta\alpha}
+ \beta^{'}\frac{\delta{\cal S}}{\delta\beta} 
+ \phi^{'}\frac{\delta{\cal S}}{\delta\phi} 
+ \xi^{'}\frac{\delta{\cal S}}{\delta\xi} - \partial_{\chi}
\left(\frac{\delta{\cal S}}{\delta\alpha}\right)\right] &=& 0.
\end{eqnarray}
The factor ordering in the momentum constraint (\ref{momentumS})
is the natural one because it implies that the wave functional depends
only on the 3-geometry of the sections $t=\mbox{constant}$. At order 
$\hbar^{0}$, Eq. (\ref{WDWS}) reduces to the Hamilton-Jacobi
(H-J) functional equation, whose solution is the phase of the WKB wave 
functional. The momentum constraint equation (\ref{momentumS}) remains 
unchanged. 
\par
As it has already been noticed, a topology change in our model can be 
studied only in a midisuperspace framework because one is forced to introduce 
a shift function in order to produce the correct Einstein's 
equations. This is technically more difficult because we now deal 
with equations which are no longer ordinary differential 
equations but functional differential equations. It is clear that 
Eqs. (\ref{WDWS}) and (\ref{momentumS}) are very complicated. We were not 
able to find solutions valid in the full midisuperspace $M$. The solutions 
we obtained are only valid in the minisuperspace ${\cal M}$, i.e., they
are solutions in the subdomain ${\cal M}\subset M$ of the full 
midisuperspace\footnote{
For instance, this is similar to obtaining solutions
of a partial differential equation, $\left[\frac{\partial}{\partial x^{2}} +
\frac{\partial}{\partial y^{2}} - \frac{\partial}{\partial t^{2}} + 
V(x,y,t)\right]f(x,y,t) = 0$, restricted to a spatial subdomain, 
say $y=0$.}. This subdomain ${\cal M}$ is defined by:
\begin{equation}
\label{midi1-2}
a^{'} = 0,\quad \phi^{'} = 0,\quad \xi^{'} = 0,\quad
\sigma = \frac{\sin\left(\sqrt{k(t)}\chi\right)}{\sqrt{k(t)}}.
\end{equation}
Under these assumptions several solutions to the complete functional 
equations (\ref{WDWS}) and (\ref{momentumS}) (i.e., in ${\cal M}$) can be 
obtained. However, they are not suitable 
for the study of topology change. We present some of them in the 
appendix. In the following  we restrict ourselves to WKB solutions. In this 
case, it is possible to find solutions where we can 
examine the possibility of a topology change that may happen 
quantum-mechanically. For $U(\phi)=V(\xi)=0$, we display two solutions 
to both the H-J and the momentum constraint functional equations
\begin{eqnarray}
\label{S1}
{\cal S}_{1} &=& \frac{1}{4\pi}\int_{V^3} {\rm d}\tau 
\biggl\{-\frac{C}{m}\bigl(\beta^{'}e^{-\alpha-\beta}\bigl)^{'}\xi  
+\beta^{'}e^{-\beta}
\Bigl[\sqrt{\frac{3-C}{2}}\phi+\sqrt{6+2C}\bigl(\alpha-\ln(-\beta^{'})\bigl)
\Bigr]\biggr\}, \\
\label{S2}
{\cal S}_{2} &=& \frac{1}{4\pi}\int_{V^3} {\rm d}\tau 
\biggl\{-\frac{C}{m}\bigl(\beta^{'}e^{-\alpha-\beta}\bigr)^{'}\xi 
+4i\sqrt{C+3}e^{\alpha - \frac{\beta}{2}}
+i\sqrt{2C+3}\beta^{'}\phi e^{-\beta}\biggr\},
\end{eqnarray}
where ${\rm d}\tau = {\rm d}\chi{\rm d}\varphi{\rm d}\theta\sin\theta$
and $C$ is an arbitrary constant. It is worth emphasizing that 
they are solutions to the momentum constraint equation in 
the full midisuperspace $M$. In order to check that these solutions 
satisfy the 
H-J functional equation, we must first perform the functional 
derivatives, then substitute them into the H-J 
equation, and finally make the restriction to 
${\cal M}$, namely, take into account the conditions (\ref{midi1-2}). Note 
that if we use the classical relations 
$\pi_{Q_{i}}={\rm \delta}{\cal S}/{\rm \delta}Q_{i}$ where the $\pi_{Q_{i}}$
are given by  Eqs. (\ref{pia}), (\ref{pisigma}), (\ref{piphi}) and 
(\ref{pichi}), we re-obtain the result $\dot{k}=0$.  
\par
The last step is to perform the integration with respect to the 
variable $\chi $. This leads to the following wave functions:  
\begin{eqnarray}
\label{Psi1}
\Psi_1 &=& \exp\bigl(\frac{i{\cal S}_1}{\hbar}\bigr) \nonumber \\ 
&=& \exp \Biggl\{ \frac{i}{4\pi\hbar}\int _0^{2\pi}{\rm d}\varphi 
\int _0^{\pi }{\rm d}\theta \sin\theta 
\biggl[\frac{C}{m}\frac{a}{\sqrt{k}}\xi \sin (2\lambda _0)                 
-\sqrt{\frac{3-C}{2}}\phi \frac{a^2}{k}\sin^2\lambda _0
\nonumber \\
& & -\sqrt{6+2C}\frac{a^2}{k}\Bigl(\sin^2 \lambda _0
\ln\bigl(\frac{a}{2\sqrt{k}}\sin \lambda _0\bigr)
+\cos \lambda _0\ln \bigl(\cos \lambda _0\bigr)\Bigr)\biggr]\Biggr\}, \\
\label{Psi2}
\Psi_{2} &=& \exp\bigl(\frac{i{\cal S}_2}{\hbar}\bigr) \nonumber \\  
&=& \exp \Biggl\{ \frac{i}{4\pi\hbar}\int_0^{2\pi}{\rm d}\varphi 
\int _0^{\pi} {\rm d}\theta \sin\theta 
\biggl[\frac{C}{m}\frac{a}{\sqrt{k}}\xi \sin(2\lambda_0)
- i\sqrt{2C+3}\phi \frac{a^2}{k}\sin^2\lambda_0 \nonumber \\
& & + 8i\sqrt{C+3}\frac{a^2}{k}\sin^2\bigl(\frac{\lambda_0}{2}\bigr)
\biggr]\Biggr\},
\end{eqnarray}
where $\lambda_{0}\equiv \sqrt{\epsilon}
\chi^{(\epsilon)}_{0}(\theta,\varphi)$ and where the symbol $\epsilon 
\equiv \mbox{sign}(k)$ is defined by $\epsilon =0$ if $k=0$ and 
$\epsilon =\pm 1$ if $k>0$ or $k<0$. The above expressions are valid 
whatever the sign of $k$ is. For instance, if $k<0$ the term 
proportional to $\xi$ becomes $(a/\sqrt{-|k|})\sin[2i\chi _0^{(-1)}(
\theta ,\varphi )]=(a/\sqrt{|k|})\sinh[2\chi _0^{(-1)}(\theta ,\varphi )]$. 
For $k=0$, this term would be equal to $2a\chi _0^{(0)}(\theta ,\varphi )$.
\par
On the other hand, under the change of coordinates 
$\bar{\chi }=\sqrt{k(t)}\chi$, 
$\bar{\theta }=\theta$, $\bar{\varphi }=\varphi$, the line element of the 
three-dimensional spacelike 
sections, see Eq. (\ref{defhypersurf}), transforms into the following 
expression:
\begin{equation}
\label{ds2change}
{\rm d}s^{2} = \frac{a^2(t)}{k(t)}[{\rm d}\bar{\chi}^2+\sin ^2(\bar{\chi })
{\rm d}\Omega^2(\theta ,\varphi)].
\end{equation}
We see that the metrics $\{k(t)>0,a(t)\}$ and $\{1,\bar{a}(t) = 
a(t)/\sqrt{k(t)}\}$ are 
simply related by a diffeomorphism. Equivalently, this 
is also true for the metrics $\{k(t)<0,a(t)\}$ and $\{-1,\bar{a}(t) = 
a(t)/\sqrt{-k(t)}\}$. In other words, they belong to the same 
equivalence class, and describe the same three-geometry in the minisuperspace, 
i.e., they 
are represented by a ``single point" in ${\cal M}$. We remark that the 
wave functions $\Psi _1$ and $\Psi _2$ depend only on the combination 
$a(t)/\sqrt{|k|}$, which can be redefined as a new scale factor $\bar{a}(t) = 
a(t)/\sqrt{|k|}$. This is exactly what was expected since $\Psi _1$ and 
$\Psi _2$ are solutions to the momentum-constraint equation and 
therefore must be invariant under three-dimensional diffeomorphisms 
on $V^3$. The wave functions $\Psi _1$ and $\Psi _2$ depend only 
on $\epsilon $ and not on $k$.
\par
Let us also emphasize that our procedure for obtaining the wave 
functions for $k=0,\pm 1$ is by no mean equivalent to the ``naive" procedure 
which would consist in constructing a minisuperspace for each value 
of $k$ and in solving the corresponding Wheeler-DeWitt equation in 
each case. Indeed, in our model the dynamical behaviour of $k(t)$ is 
encoded in the momentum $\pi _{\sigma }$ which does not appear at 
all in the naive approach where $k$ is fixed. It is well-known that in 
this case the gravitational sector of the Wheeler-De Witt equation contains 
only $\pi _a$. It is easy to see that the presence of this term 
modifies drastically the problem. The naive approach is therefore 
completely different from the formalism considered here.
\par
The above wave functions $\Psi_{1}$ and $\Psi_{2}$ are the main result
of this section. In the next paragraph, we will try to extract physical 
information from them.

\section{Interpretation}
\subsection{Standard Probabilistic Interpretation}

In this section, we turn to the problem of the interpretation of the 
wave function. Conceptually, this question is one of the most difficult 
challenges one faces in trying to construct a consistent theory of 
Quantum Gravity. In 
General Relativity, the principle of covariance is expressed through the 
diffeomorphism invariance which leads to the constraints. After 
quantization {\it \`a la Dirac}, those constraints transform into 
functional differential equations for the state $\Psi $. The momentum 
constraint indicates that $\Psi $ depends only on the three-geometry of 
the spacelike sections and therefore 
is a functional on superspace. The Hamiltonian constraint expresses the 
fact that $\Psi $ is independent of the time variable occurring in the 
Hamiltonian formalism. This fact is understandable since $\Psi $ is supposed 
to describe the Universe as a whole and therefore, by definition, cannot 
depend on some exterior parameter. This causes severe problems when one 
tries to interpret Quantum Gravity because time plays a fundamental role 
in the structure of Quantum Mechanics \cite{Unhru}. 
\par
The situation described above seems to be similar to what happens in the 
parametrized version of non-relativistic particle mechanics. In this theory, 
the path followed by the particle can be parametrized by a label $\tau $, and 
the absolute Newtonian time $t$ can be considered as one of the dynamical 
variables. After quantization, this approach leads to an equation $H\Psi =0$ 
comparable to the Wheeler-DeWitt equation. However, in this case, there 
exists a criterion which allows one to identify immediately what is the 
time variable: the 
Hamiltonian is linear in the momentum canonically conjugated to $t$, whereas it 
is quadratic in the momenta conjugated to the space variables. Hence, the 
equation $H\Psi =0$ reduces to the ordinary Schr\"odinger equation.
\par
In Quantum Gravity the situation is different. The Hamiltonian constraint, 
i.e., the Wheeler-DeWitt equation, is quadratic in all the momenta, and a 
priori we have no mean to recognize what is time in this context. However, 
the idea to introduce a Hamiltonian which is linear in one of the momenta 
has been pursued by many authors. The extrinsic time approach advocated by 
Teitelboim and Kucha$\check{\mbox{r}}$ \cite{tei,kuc2} is one of these proposals. It 
consists in performing a canonical transformation such that the Wheeler-De 
Witt Hamiltonian becomes linear in the momentum conjugated to the trace of the 
extrinsic curvature $K$ (which should be viewed as time). However, 
the corresponding Schr\"odinger Hamiltonian is known only through the 
solution of a complicated differential equation. Therefore, this approach 
remains unsatisfactory.
\par
Another proposal has been made by Kucha$\check{\mbox{r}}$ and Torre 
\cite{21}. The idea is to introduce 
a dust field $\xi (t,x^i)$ as one of the sources of matter because the 
Hamiltonian is first order in $\pi _{\xi}$. This was our main motivation for 
having considered $\xi (t,x^i)$ in the previous sections. Even if the 
interpretation of this dust field in the very early Universe remains rather 
obscure, it gives us a well-defined structure where we can ask physical 
relevant questions. As a matter of fact, limitations will come only from 
technical problems. 
\par
In a first step, we will try to use this framework in order to interpret 
the solutions of section 3, and to see if they describe a spacetime which 
displays a change of topology. 
\par
It is clear that the most efficient method would be to compute the quantum 
propagator defined by:
\begin{equation}
\label{propa}
K(a_i,\sigma _i,\phi _i,\xi _i|a_f,\sigma _f,\phi _f,\xi _f)=
\int {\cal D}a{\cal D}\sigma {\cal D}\phi\,\mu(a,\sigma,\phi) 
e^{\frac{i}{\hbar}S[a,k,\phi,\xi]}.
\end{equation}
where $\mu(a,\sigma,\phi)$ is the measure.
Then, expressing this propagator for the function $\sigma (t, \chi )$ given 
in Eq. (\ref{midi1-2}) we would be able to determine the probability of having 
$k_f=1$, for instance, at ``time" $\xi _f$ knowing that $k_i$ was equal to 
$-1$ (or $0$) at initial time $\xi _i$. A non vanishing result would 
demonstrate explicitly the possibility of a change of topology at the 
quantum level. Unfortunately, computing the propagator (\ref{propa}) is 
technically very complicated. Another possibility would be to evaluate 
$K(a_i,\sigma _i,\phi _i,\xi _i|a_f,\sigma _f,\phi _f,\xi _f)$ at the 
semi-classical level using the well-known result of Van Vleck and Gutzwiller
\cite{Van Vleck,Gutzwiller}. 
However, this would require the knowledge of all the classical trajectories.
\par
Having only at our disposal the states given by Eqs. (\ref{Psi1}) and 
(\ref{Psi2}), we can also try to compute the mean value of different 
interesting quantities. For example, a good candidate would be the 
three-curvature:
\begin{equation}
\label{meanR}
<\mbox{}^{3}\hspace{-.1cm}R>\equiv <\Psi|\mbox{}^{3}\hspace{-.1cm}R|\Psi>.
\end{equation}
Another natural possibility would be the operator $-\sigma ''/\sigma $ 
which, in the minisuperspace ${\cal M}$ where 
the solutions (\ref{Psi1})-(\ref{Psi2}) are valid, reduces simply to $k$:
\begin{equation}
\label{meank}
<k>\equiv \left.<\Psi|-\frac{\sigma ''}{\sigma }|\Psi>\right|_{{\cal M}}.
\end{equation}
In what follows, we will restrict our considerations to the calculation of 
Eq. (\ref{meank}) in ${\cal M}$. Then, we will be able to study the sign of 
$k$ as a function of $\xi $, the ``time" in our system. 
\par
It has been emphasized above that, in order to compute mean values and norm 
of wave functionals, a measure should be chosen. In the previous section, we 
adopted the cartesian measure ${\cal D}\alpha {\cal D}\beta {\cal D}\phi $. 
Therefore, after having put the Wheeler-DeWitt equation in the form of a 
Schr\"odinger equation, the solutions should be normalized according to the 
formula:
\begin{equation}
\label{defnorme}
<\Psi|\Psi> = \int _{M}{\cal D}\alpha {\cal D}\beta {\cal D}\phi 
\Psi [\alpha,\beta,\phi,\xi]\Psi ^*[\alpha,\beta,\phi,\xi].
\end{equation}
The fact that the Hamiltonian, with the cartesian measure, is hermitian 
automatically guarantees that $<\Psi|\Psi>$ will not depend on $\xi $. In 
practice computing the integral (\ref{defnorme}) is a difficult task. In 
addition, this requires the knowledge of the wave functional in the full 
midisuperspace $M$, whereas we know $\Psi $ only in the minisuperspace 
${\cal M}\subset M$. Endowed with these solutions, we can only write a 
necessary but not sufficient condition in order to have a normalized wave 
functional, namely:
\begin{equation}
\label{normeinM}
<\Psi|\Psi>_{{\cal M}} \equiv \int _{{\cal M}\subset M}{\cal D}\alpha 
{\cal D}\beta 
{\cal D}\phi \Psi [\alpha,\beta,\phi,\xi] \Psi ^*[\alpha,\beta,\phi,\xi] 
<\infty.
\end{equation}
This equation can be worked out.  
The ranges of $\phi $ 
and $\xi $ are given by $ -\infty <\phi <+\infty $ and 
$-\infty <\xi <+\infty$. The domain of $a$ is $(0,+\infty )$, and from 
the discussion of the previous section we see that integrating over $k$ 
will amount to sum over $k\in\{0,\pm 1\}$. The result is that 
$<\Psi |\Psi>_{{\cal M}}$ can be put under the form:
\begin{eqnarray}
\label{reduction1}
<\Psi|\Psi>_{{\cal M}} &=& 
\int _{-\infty }^{\infty } {\rm d}\alpha \sum _{k=0,\pm 1}
\int _{-\infty }^{\infty }{\rm d}\phi \int _{M}{\cal D}\alpha 
{\cal D}\beta {\cal D}\phi\, \delta [\alpha (t,\chi)-\alpha (t)] 
\nonumber \\
& & \delta \left\{\beta (t,\chi) + \ln\left[\frac{k(t)}{a^{2}\sin^{2}
(\sqrt{k(t)}\chi)}\right]\right\} 
\delta [\phi(t,\chi )-\phi(t)] \nonumber \\
& & \Psi [\alpha,\beta,\phi,\xi] \Psi ^*[\alpha,\beta,\phi,\xi] \\
\label{reduction2}
&=& \sum _{k=0,\pm 1}\int _{-\infty }^{\infty } {\rm d}\alpha 
\int _{-\infty }^{\infty }{\rm d}\phi 
\Psi (\alpha,k,\phi,\xi)\Psi ^*(\alpha,k,\phi,\xi),
\end{eqnarray}
where $\delta [f-g]$ is the $\delta $-functional defined by the expression 
(see Ref. \cite{Hatfield}):
\begin{equation}
\label{defdelta}
\delta [f-g] \equiv \prod _{\chi }\delta (f(\chi )-g(\chi )).
\end{equation}
Finally, since $\alpha \equiv \ln a$, we obtain the following condition:
\begin{equation}
\label{finalred}
<\Psi|\Psi>_{{\cal M}} =\sum _{k=0,\pm 1}\int _0^{+\infty } 
\frac{{\rm d}a}{a}\int _{-\infty }^{\infty }{\rm d}\phi \Psi (a,k,\phi,\xi)
\Psi ^*(a,k,\phi,\xi)<\infty .
\end{equation}
Let us now come back to our solutions $\Psi _1$ and $\Psi _2$, which we will 
denote by $\Psi _i$, $i=1,2$. They depend on an arbitrary complex 
constant $C$. Therefore, the principle of superposition implies that 
the most general state satisfying the Wheeler-DeWitt equation which we can 
construct with a single state $\Psi _i$, is given by:
\begin{equation}
\label{genestate}
\Psi _i(a,k,\phi,\xi)=\int _{{\cal C}}{\rm d}Cf_i(C)e^{\frac{i}{\hbar }
S_i(C;a,k,\phi,\xi)},
\end{equation}
where $f_i(C)$ is an arbitrary function. Of course, we could also 
construct wave functions which would be superpositions of the states 
$\Psi _1$ and $\Psi _2$. The function 
$f_i(C)$ should be chosen such that $\Psi $ be normalized. 
Let us fix the $S^{3}$ topology for $k=1$. It follows that 
$0\leq\chi\leq \pi$ and $S_i(C;a,k=1,\phi,\xi)=0$ for any value of $C$ 
[see Eqs. (\ref{Psi1}) 
and (\ref{Psi2})]. Hence, the states $\Psi _i$ reduces to:
\begin{equation}
\label{statek=1}
\Psi _i(a,k=1,\phi,\xi)=\int _{{\cal M}}{\rm d}Cf_i(C).
\end{equation}
Therefore, in the sum over $k$ in Eq. (\ref{finalred}), the contribution 
coming from the term $k=1$ is equal to:
\begin{equation}
\label{k=1term}
\int {\rm d}C{\rm d}C'f_i(C)f_i^*(C')\int _0^{+\infty } \frac{{\rm d}a}{a}
\int _{-\infty }^{\infty }{\rm d}\phi ,
\end{equation}
and the only way to obtain a normalized wave function is to choose $f_i(C)$ 
such that:
\begin{equation}
\label{conditionf}
\left|\int {\rm d}C f_i(C)\right|^{2} = 0.
\end{equation}
{}From Eq. (\ref{statek=1}) we see that $\Psi_{i}(k=1)=0$.
Note that this condition is a necessary condition in order to obtain a 
normalized wave functional extended from the solutions $\Psi_{1}$ and 
$\Psi_{2}$
to the full midisuperspace $M$. Of course, $f_i(C)$ must also be taken such 
that the two other contributions coming from the terms $k=0$ and $k=-1$ in the 
sum (\ref{finalred}) be finite.
\par
Let us now turn to the computation of the mean value of $k$. 
By definition, the expression for $<k>$ can be written as:
\begin{eqnarray}
\label{resultmeank}
<k>(\xi) &=& \sum _{k=0,\pm 1}\int _0^{+\infty } \frac{{\rm d}a}{a}
\int _{-\infty }^{\infty }{\rm d}\phi \Psi (a,k,\phi,\xi)k
\Psi ^*(a,k,\phi,\xi) \\
\label{signk}
&=& -\int _0^{+\infty } \frac{{\rm d}a}{a}\int _{-\infty }^{\infty }{\rm d}
\phi \Psi (a,k=-1,\phi,\xi)\Psi ^*(a,k=-1,\phi,\xi),
\end{eqnarray}
where the last formula was obtained using the property (\ref{conditionf}). 
{}From Eq. (\ref{signk}) we deduce that:
\begin{equation}
\label{selectionrule}
<k>(\xi)\le 0.
\end{equation}
We remark that this result is rigorous in the sense that it has been 
established from considerations valid in the full midisuperspace $M$. 
Equation (\ref{selectionrule}) tells us that the change of topology 
from $S^{3}$ to $k=0$ or $k=-1$ is forbidden whereas the passage from 
$k=0$ to $k=-1$, or the opposite\footnote{It would be interesting to 
know if it is more likely to have a transition from $k=0$ to $k=-1$ or 
from $k=-1$ to $k=0$. In the framework developed here, this information
can not be given}, is {\it a priori} allowed. Of course 
this ``selection rule" is valid only for the solutions presented in 
section 3, assuming the $S^{3}$ topology for $k=1$, and
 at the $\hbar ^0$ order. It is interesting to note that,
already at this level, the different possibilities are not equivalent. 
In order to demonstrate that a change of topology from $k=0$ to $k=-1$ 
is actually possible, an 
explicit function $f_i(C)$ should be considered. Also, if we had chosen
the $D^{3}$ topology for $k=1$, then 
$0\leq\chi\leq\chi_{0}^{(1)}(\theta,\varphi)$
and $S_{i}(C;a,k=1,\phi,\xi) $ would no longer be a vanishing 
quantity. Hence, it would no longer be necessary to
have $\Psi_{i}(k=1) = 0$ in order to have a normalizable wave 
function. In principle, topology change from $k=1$ and a topology 
different from $S^3$ could be possible in the
standard probabilistic interpretation. However, we leave these 
questions open and turn now to a different scheme of interpretation.

\subsection{The Conditional Probability Interpretation}

The result of the previous section crucially depends on our ability to 
normalize the wave function. However, it has been recognized long time 
ago that unnormalized wave functions can also be used to make 
predictions if one 
utilizes them to compute conditional probabilities. This approach has been 
advocated by many authors in the context of minisuperspaces, where the wave 
function is in general not normalizable. In the case of a FLRW spacetime 
filled 
with a scalar field, this allows one to compute the probability distribution 
for the initial value of the scalar field \cite{Halliwell}. Here, we will use 
this formalism in order to learn something about $k$. 
\par
Suppose we have found that a spacelike hypersurface is homogeneous and 
isotropic, with the scale factor given by some value $a=\bar{a}$, and the
scalar field given by $\phi=\bar{\phi}$. Then, the
conditional probability of having this spacelike hypersurface with
$k$ equal to $-1$, $0$ or $+1$, at some ``time" $\xi$, 
knowing that $a=\bar{a}$ and $\phi=\bar{\phi}$, will be given by:
\begin{equation}
\label{defPc}
P_c(k|\bar{a},\bar{\phi}) \equiv \frac{|\Psi (k,\bar{a},\bar{\phi})|^2}
{\sum _{k=0,\pm 1}|\Psi (k,\bar{a},\bar{\phi})|^2}.
\end{equation}
The constraint $\sum _{k=0,\pm 1}P_c(k|\bar{a},\bar{\phi})=1$ can be 
immediately checked. Let us 
compute and discuss $P_c(k|\bar{a},\bar{\phi})$ for the wave 
function $\Psi _2$. At this stage, 
boundary conditions 
on $\Psi _2$ should be chosen. In the present context, this amounts to fix 
the value of the constant $C$ [for the solution (\ref{genestate}) this would 
amount to choose the function $f_i(C)$ and the contour ${\cal C}$]. Many 
proposals have been made to answer this 
question \cite{HH,Vilenkin}. We will consider two different possible 
values for $C$ in order to illustrate 
that different choices actually lead to different predictions. Here, our 
aim is 
just to exhibit at least one state for 
which the change of topology considered in this article is possible, 
and it should be clear that the two wave functions $\Psi _1$ and $\Psi _2$ 
do not represent automatically states of cosmological interest. 
\par
As a first example, let us assume that $C=-i$. The term multiplying 
$\xi$ in the exponential of Eq. (\ref{Psi2}) becomes:
\begin{eqnarray}
\label{F1}
F_{1}(V^3) &\equiv & \frac{\bar{a}}{4\pi\hbar m}
\int_{0}^{2\pi} {\rm d}\varphi \int_{0}^{\pi} {\rm d}\theta
\sin\theta
\sin\left[2\chi_{0}^{(1)}(\theta,\varphi;V^3)\right],
\quad {\mbox if} \quad k=1,  \\
\label{F0}
F_{0}(T^3) &\equiv & \frac{\bar{a}}{4\pi\hbar m}
\int_{0}^{2\pi} {\rm d}\varphi \int_{0}^{\pi} {\rm d}\theta
\sin\theta
[2\chi^{(0)}_{0}(\theta,\varphi)], 
\quad {\mbox if} \quad k=0, \\
\label{F-1}
F_{-1}(I^3) &\equiv & \frac{\bar{a}}{4\pi\hbar m}
\int_{0}^{2\pi} {\rm d}\varphi \int_{0}^{\pi} {\rm d}\theta
\sin\theta
\sinh\left[2\chi_{0}^{(-1)}(\theta,\varphi)\right],
\quad {\mbox if} \quad k=-1.
\end{eqnarray} 
The numerical values on the $F_i$'s determine 
the behaviour of the conditional probabilities. If $V^3=S^3$, then 
$F_1(S^3)=0$. For the other cases it is not possible to calculate the 
$F_i$'s exactly because we do not know the form of $\chi _0^{(\epsilon)}
(\theta,\varphi)$. But we can put bounds on their numerical values, and it 
will be sufficient for our purpose here. Let us start with $F_1(D^3)$. 
We know (see the Introduction) that $\chi ^{(1)}_{max}=0.163$. For the 
interval of integration under consideration, the integrand in  
Eq. (\ref{F1}) is positive. This implies that:
\begin{equation}
\label{borneF1}
F_1(D^3) \leq \frac{\bar{a}}{\hbar m}\sin [0.326]\approx 
0.320\frac{\bar{a}}{\hbar m}.
\end{equation}
In the same manner, when $k$ is negative, we can establish that:
\begin{equation}
\label{borneF-1}
2.75 \frac{\bar{a}}{\hbar m}\leq F_{-1}(I^3) \leq 
7.90 \frac{\bar{a}}{\hbar m}.
\end{equation}
Note that $F_{-1}$ is always greater than $F_1$. The case $k=0$ is 
slighty different. As pointed out before, the value of $\chi ^{(0)}_{max}$ 
depends on the fundamental cube of $T^3$, which is arbitrary. Hence, $F_0$ 
is also arbitrary.
\par
Using the definition (\ref{defPc}) and Eq. (\ref{Psi2}), we obtain the 
following expressions for the conditional probabilities:
\begin{eqnarray}
\label{valuePc1}
P_c(k=-1) &=& \frac{A(\bar{a},\bar{\phi})e^{2F_{-1}\xi}}
{A(\bar{a},\bar{\phi})e^{2F_{-1}\xi} 
+ B(\bar{a},\bar{\phi})e^{2F_{0}\xi} 
+ C(\bar{a},\bar{\phi})e^{2F_{1}\xi}}, \\
\label{valuePc2}
P_c(k=0) &=& \frac{B(\bar{a},\bar{\phi})e^{2F_{0}\xi}}
{A(\bar{a},\bar{\phi})e^{2F_{-1}\xi} 
+ B(\bar{a},\bar{\phi})e^{2F_{0}\xi} 
+ C(\bar{a},\bar{\phi})e^{2F_{1}\xi}}, \\
\label{valuePc3}
P_c(k=+1) &=& \frac{C(\bar{a},\bar{\phi})e^{2F_{1}\xi}}
{A(\bar{a},\bar{\phi})e^{2F_{-1}\xi} 
+ B(\bar{a},\bar{\phi})e^{2F_{0}\xi} 
+ C(\bar{a},\bar{\phi})e^{2F_{1}\xi}},
\end{eqnarray}
where the positive functions $A(\bar{a},\bar{\phi} )$, 
$B(\bar{a},\bar{\phi} )$ and $C(\bar{a},\bar{\phi})$ can be calculated 
from Eq. (\ref{Psi2}). Since we will be interested in the behaviour 
of $P_c(k)$ at the boundaries of the minisuperspace in $\xi$, these
functions will not be relevant for what follows because they do not
depend on $\xi$. For clarity, we will treat the two possibilities for 
$k=1$ considered in this article, separately.
\par
If $V^3=S^3$, then $F_{1}=0$ and we have two possibilities: $F_{0}>F_{-1}$
or $F_{0} < F_{-1}$. From the previous expressions, we get:
\begin{equation}
\label{limit-1,0,1}
\lim _{\xi \to -\infty }P_c(k=-1)=\lim _{\xi \to -\infty }P_c(k=0)=0,
\quad {\mbox and} \quad \lim _{\xi \to -\infty }P_c(k=+1)=1.
\end{equation}
On the other hand, if we compute the same probabilities but this time when 
the dust field goes to $+\infty $, we obtain:
\begin{eqnarray}
\label{limit2}
\lim _{\xi \to +\infty }P_c(k=+1) &=& 0, \\
\label{limit-1}
\lim _{\xi \to +\infty }P_c(k=-1) &=& 
\left\{ \begin{array}{ll} 
0 & \quad {\mbox if} \quad F_{0} > F_{-1}, \\
1 & \quad {\mbox if} \quad F_{0} < F_{-1},
\end{array}
\right. \\ 
\label{limit0}
\lim _{\xi \to +\infty }P_c(k=0) &=& 
\left\{ \begin{array}{ll} 
0 & \quad {\mbox if} \quad F_{0} < F_{-1}, \\
1 & \quad {\mbox if} \quad F_{0} > F_{-1}.
\end{array}
\right.  
\end{eqnarray}
Therefore, in this case, definite predictions can be made since conditional 
probabilities are either equal to $0$ or $1$. If $F_{0} < F_{-1}$
there is a change of topology from $k=1$ to $k=-1$ when we go from 
$\xi =-\infty $ to $\xi =+\infty $ in the minisuperspace. On the other 
hand, if $F_0>F_{-1}$, there is a change of topology from 
$k=1$ to $k=0$. It is clear that this is a quantum mechanical effect since 
it is not possible that a classical path could connect the two regions. A 
similar result 
would have been obtained if we had used $\Psi _1$ instead of $\Psi _2$.
\par
For the set $\{D^{3},T^{3},I^{3}\}$, we have $0 < F_{1} < F_{-1} $ and
three possibilities:\\
a) $F_{0} < F_{1}$. The conditional probabilities 
(\ref{valuePc1})-(\ref{valuePc3}) give for $\xi \to - \infty$ 
$$     
P_c(k=0)= 1
$$
and for $\xi \to +\infty$
$$
P_c(k=-1)= 1;
$$
b) $F_{1}<F_{0}<F_{-1}$. For $\xi\to -\infty$ we have
$$
P_c(k=+1)= 1
$$
and for $\xi\to +\infty$,
$$
P_c(k=-1)= 1;
$$
c) $F_{1}<F_{-1}<F_{0}$. For $\xi\to -\infty$ we have
$$
P_c(k=+1)= 1
$$
and for $\xi\to +\infty$,
$$
P_c(k=0)= 1.
$$

It is interesting to notice that the behaviour of the conditional 
probabilities for the choice $C=i$ can be obtained directly from the 
previous results (obtained for $C=-i$) just by replacing 
$\xi $ by $-\xi$. This shows the importance of the choice 
of boundary conditions on the final result.
\par
At this point, we would like to make an important remark. The conditional 
probabilities exhibited before possess an interesting property. It 
seems that it exists ``privileged directions" for the change of 
topology. For $C=-i$, for instance, 
it always occurs from the smallest $F_i$ to the biggest 
$F_j$. Note that the $F_{i}$ are as big as it is the volume of the
closed hypersurfaces they are referred to. 
Whether this property is generic or 
just a consequence of the particular approximated solutions 
found here is, of course, a question which remains to be solved.
\par
Let us conclude this section by emphasizing that different 
interpretations actually lead to very different results. For instance, the 
``selection rule" established previously is violated in the conditional 
probability interpretation. This shows how crucial the problem of the 
interpretation of Quantum Gravity is.

\section{Conclusion}

In this article, we have addressed the problem of topology change in 
canonical Quantum Cosmology. More precisely, we have treated the question 
of whether a quantum change of topology  can occur between two 
spacelike hypersurfaces of a FLRW metric whose intrinsic curvatures have 
a different sign.
\par
A consistent
Hamiltonian treatment of this problem requires an enlargement
of the {\it minisuperspace} model to a {\it midisuperspace} model. This 
enlarged 
configuration space is, {\it per se}, very interesting to analyze
because the corresponding theory contains an infinite number of degrees 
of freedom and is therefore a (Quantum) Field Theory, closer to the 
full Quantum Gravity
than the usual minisuperspace models. New problems arise, like
anomalies and regularization, and the Wheeler-DeWitt equation becomes
a functional differential equation. We have also seen that the 
physical predictions depend crucially on the
interpretation we adopt. In the standard probabilistic 
interpretation, we have shown that the changes of topology 
$S^3(k=+1)\rightarrow T^3(k=0)$ or $S^3(k=+1)\rightarrow I^3(k=-1)$ are 
forbidden for the particular set of solutions we have obtained. 
On the other hand, if we apply the conditional probability 
interpretation to this same set of solutions,  a wide class of quantum 
changes of topology are actually possible, including the ones forbidden
by the standard probabilistic interpretation mentioned above. 
\par
There are many perspectives for future works. One could try to find
an exact normalizable solution (with the measure defined in section 3) 
in the full midisuperspace $M$, and try to use the probabilistic
interpretation in order to check if topology changes can occur in 
this case. In 
fact, exact solutions to the Hamilton-Jacobi equations (\ref{WDWS}) and 
(\ref{momentumS})
already exist in the literature but only with $\phi=\xi=0$ \cite{Kief} 
or with
$\xi =0$ and a dilaton field \cite{Kuc}. It should be very interesting
to study if these solutions can predict a topology change\footnote{Also, 
it would be worthwhile to examine if the predictions
given by these midisuperspace solutions agree with the ones known in the
literature for the minisuperspace models they contain, 
see Ref. \cite{Kucharcon}.}. New 
families of solutions involving only the more fundamental scalar field 
could also be obtained by mean of the the long-wavelength approximation 
\cite{sal1}. In this case, topology changes should be examined between 
hypersurfaces characterized by different values of the scale factor. 
\par
In this paper, we have restricted ourselves to the study of the spaces 
$\{S^3,D^3,T^3,I^3\}$. As pointed out in the Introduction, many other 
cases are {\it a priori} possible. Therefore, one could calculate 
the corresponding conditional probabilities in order to 
see if new topological transitions are possible. Note that even open spaces
with $k=0$ could be also considered because the 
surface terms which could appear in the Hamiltonian formalism vanish 
in this case.
\par
It would be interesting to calculate the probability amplitude for having
change of topology of the surfaces of homogeneity by using the path integral 
approach. To our knowledge, this approach has never been applied to this
particular problem. In principle one should find a solution of the classical
equations in the Euclidean midisuperspace which, when reduced to the 
minisuperspace subspace, could be interpreted as an instanton connecting two
homogeneous hypersurfaces with distinct topologies.  The exponential of the
Euclidean action evaluated at this solution would then give the semi-classical
probability amplitude for change of topology. To find such a solution is,
however, a very difficult task.
\par
We hope to address these questions in future publications.

\section*{Acknowledgments}
We are grateful to G. Cl\'ement, H. Fagundes, L. Ford, R. Kerner,
B. Linet, V. Moncrief, P. Spindel, A. F. da F. Teixeira and the
group of the ``Pequeno Semin\'ario'' for useful discussions. R. Portugal
is acknowledged by providing us with his maple program ``RIEMANN''. 
We thank CNPq for financial support.

\section*{Appendix}

In this appendix, we present solutions to the Wheeler-DeWitt and momentum 
constraint equations which are different from those already found in 
Eqs. (\ref{S1}) and (\ref{S2}). All of them are solutions restricted to 
the minisuperspace ${\cal M}$. Although they are of limited 
interest for the problem of topology change, they could be compared 
to the solutions directly obtained from the minisuperspace 
formulation. In the variables ($\alpha,\beta$) they can be written as:
\begin{equation}
\Psi = \exp \Bigl(\frac{i}{\hbar}{\cal S}[\alpha,\beta,\xi,\phi]\Bigr).
\end{equation}
\par
A first class of solutions is valid only for the case 
${}^3\hspace{-.1cm}R=0$. Obviously, they do not allow a change of 
topology. An example is given by:
\begin{equation}
{\cal S}[\alpha,\beta,\phi] =\frac{1}{4\pi} \int _{V^3}{\rm d}\tau 
\left[ e^{\alpha - \beta} e^{-\frac{9\beta}{4} \pm \sqrt{3}\phi}\right].
\end{equation}

These wave functionals are solutions to the momentum constraint 
equation (\ref{momentumS}) and the full Wheeler-DeWitt equation 
(\ref{WDWS}) with ${}^3\hspace{-.1cm}R=0$.
\par
A second class of solutions solves the functional equations (\ref{WDWS}) and 
(\ref{momentumS}) for ${}^3\hspace{-.1cm}R\ne 0$. They are given by:
\begin{eqnarray}
\label{antlast}
{\cal S}[\alpha,\beta,\xi,\phi] &=&\frac{1}{4\pi} \int _{V^3}{\rm d}\tau\Bigl[-3\bigl(\beta^{'}
e^{-\alpha-\beta}\bigr)^{'}\xi+ Ce^{\alpha - \frac{\beta}{4}}
+ \sqrt{3}\beta^{'}\phi e^{-\beta}\Bigr], \\
\label{last}
{\cal S}[\alpha,\beta,\phi] &=& \frac{1}{4\pi}\int _{V^3}{\rm d}\tau \beta^{'}e^{-\beta}
\Bigl[\sqrt{\frac{3}{2}}\phi
+ \sqrt{6}\bigl(\alpha-\ln(-\beta^{'})\bigr)\Bigr],
\end{eqnarray}
where $C$ is an arbitrary constant. We note that, for both Eqs. 
(\ref{antlast}) and (\ref{last}), the 
corresponding wave functions lead to ``time-independent" conditional 
probabilities.
\par
We would like to remark that the last functional (\ref{last}) is an
{\it exact} solution to the Wheeler-DeWitt functional equation
restricted to ${\cal M}$, while (\ref{antlast}) is a WKB solution to
Eq. (\ref{WDWS}).

\end{document}